# Predictive Strategies for the Control of Complex Motor Skills: Recent Insights into Individual and Joint Actions

Marta Russo[1], Antonella Maselli[1,2], Dagmar Sternad*[3], Giovanni Pezzulo*[1]

[1] *Institute of Cognitive Sciences and Technologies (ISTC), National Research Council (CNR), Via Giandomenico Romagnosi 18A, 00196 Rome, Italy*

[2] *Department of Biomedical and Dental Sciences and Morphofunctional Imaging, University of Messina, Via Consolare Valeria, 1, 98124 Messina, Italy*

[3] *Institute for Experiential Robotics, Department of Biology, Department of Electrical and Computer Engineering, Department of Physics, Northeastern University, Boston, Massachusetts, United States of America*

* These authors contributed equally.

**Abstract**

Humans perform exquisite sensorimotor skills, both individually and in teams, from athletes performing rhythmic gymnastics to everyday tasks like carrying a cup of coffee. The “predictive brain” framework suggests that mastering these skills relies on predictive mechanisms, raising the question of how we deploy predictions for real-time control and coordination. This review highlights two research lines, showing that during the control of complex objects people make the interaction with ‘tools’ predictable; and that during dyadic coordination people make their behavior predictable and legible for their partners. These studies demonstrate that to achieve sophisticated motor skills, we play “prediction tricks”: we select subspaces of predictable solutions and make sensorimotor interactions more predictable and legible by and for others. This synthesis underscores the critical role of predictability in optimizing control strategies across contexts. Furthermore, it emphasizes the need for novel studies on the scope and limits of predictive mechanisms in motor control.

**Highlights**

- Humans use "prediction tricks" to simplify complex motor tasks.
- Increased task predictability enhances control by preempting errors.
- Motor skills rely on selecting subspaces of predictable solution for efficiency.
- Humans interpret co-actors' kinematics, while co-actors make their actions clearer.

## Introduction

Humans master sophisticated physical skills, both individually and in teams, as evident by watching athletes performing gymnastics, volleyball or soccer. But also seemingly mundane tasks like transporting a cup filled with coffee pose unforeseen challenges. How we coordinate our high-dimensional body in interaction with complex objects, such as rings, ribbons, or a cup of coffee, is a fundamental and largely unsolved question in motor control and neuroscience. Any insights into these complex control problems will have immediate applications in sport science, rehabilitation, and robotics.

Formal principles from cybernetics, control theory, information theory, nonlinear dynamics, and, more recently, from the active inference approach can help achieve a principled understanding of how we control our actions and interactions. For example, a key idea from the cybernetic tradition is that complex behaviors could be structured into a hierarchy, in which higher levels specify targets and goals, e.g., a specific body configuration in gymnastics, and lower levels guide the execution to achieve them, implementing a closed-loop "control of perception" [1,2]. The active inference framework suggests that these motor control loops are implemented by first generating predictions about intended movement outcomes (proprioceptive and exteroceptive) and then control to minimize prediction errors, i.e., discrepancies between actual sensations and the predicted outcomes [3–5]. Active inference reconciles closed-loop control with the idea of a "predictive brain": a brain that acts as a statistical organ and learns a generative model of the world and of the body in it. The brain uses this model to continuously generate predictions that serve as targets, guide perception (predictive coding) and control actions (active inference) [6–8].

Generative modeling and predictive coding assume the presence of statistical regularities in the environment and in body-environment interactions that can be exploited for prediction and control. However, applying these ideas to sophisticated motor skills poses a fundamental challenge: the dynamics of body and tool is not easy to predict – or is not even predictable. This becomes apparent when handling tools that present high-dimensional nonlinear dynamics (e.g., shoelaces or a whip) [9], when batting a fast moving ball [10], when performing dyadic actions that require predicting others' actions, such as playing basketball [11], soccer [12], throw-and-catch games [13,14], and during parent-child play [15]. These skills are challenging to learn for a brain that seeks statistical regularities for generative and predictive models – and in which accurate control requires high-quality predictions.

One possibility to master this challenge comes from the fact that virtually all motor behaviors have redundancy, i.e., there is not only a single solution, but a manifold of solutions, whose predictability is likely to vary. This raises the hypothesis that learning such skills relies on exploring task solutions for predictable and thereby controllable subspaces. Importantly, the statistical regularities in motor control tasks not only reside in the external world, but are produced by one's own actions and, during collaborative tasks, by the partner's actions. This leads to the follow-up hypothesis that people might act in ways that select and/or create exploitable statistical regularities in the action-to-outcome mapping, rendering complex sensorimotor tasks more predictable and achievable. In this paper, we discuss two sets of studies illustrating how people exploit such "prediction tricks" during individual motor control and joint action between partners.

Importantly, there are at least two ways to understand the notion of predictability, in predictive versus non-predictive accounts of motor control. Consider the case of tracking a moving target with your gaze. Predictive theories, such as optimal motor control [16] and active inference [3,17,18], argue that the brain generates predictions about the situation (e.g. the trajectories of the target and/or the end effectors, such as head and eyes) and uses them to guide the action (e.g., move the head and eyes to fixate the target). Non-predictive theories, such as perceptual control theory [1,2,19], argue that what appears to be prediction by the nervous system are in fact emergent properties of closed loop hierarchical perceptual control: controlled variables (e.g. the perceived velocity of a moving target on the retina) and their reference values (e.g. to keep this velocity at zero) are specified rather than predicted. In these accounts, once the agent's own behavior is allowed to control the perceptual variable, prediction may become unnecessary. Therefore, the fact that something we measure is predictable, e.g., the target trajectory, does not necessarily imply that the brain predicts it. We return to this distinction at the end of the manuscript, but it's important to keep it in mind when evaluating the body of literature that we review below.

**Predictability in individual motor control**

Accurate motor control requires continuous monitoring of whether and how the executed actions approach the set goal. Over the last three decades, motor control studies employed visuo-motor rotation and force-field paradigms to investigate adaptation strategies [20]. These scenarios present examples for closed-loop control as agents receive feedback about their actions, e.g., the perceived relation between a hand and a visible target during reaching, at every instant [21].

In this context, the framework of optimal feedback control has been applied to sensorimotor control [16]. Optimal feedback control aims to minimize a cost by balancing between feedforward and feedback control along a continuum that depends on the extent to which the estimate of current body state is influenced by predictions (feedforward) or by sensory input (feedback). However, this framework poses some challenges: first and foremost, to overcome sensory delay the feedforward part of the controller needs to predict [22]. This might be more evident in tasks in which predicting objects motion is fundamental, i.e., in catching. This set of tasks not only require predicting the agent's state, but also the object's dynamics, as the visual system cannot track the object in real-time due to sensorimotor delays; it therefore needs to predict its position [23].

In addition, returning a fast ball in tennis, or even catching a thrown ball after a full turn as in gymnastics, offers less options for continuous feedback. For this reason, several scientists suggested that in addition to closed-loop tasks successful interactions with the dynamic environment also benefit from predictions – although, as noted in the Introduction, this is still debated [24]. For example, interceptive actions that involve hitting or avoiding an object in motion involve explicit or implicit predictions about the object's trajectory, augmented by on-line visual information [23]. Furthermore, children learn to predict the motion of objects as they develop, increasing their ability to catch balls [25]. When this ability is impaired, as seen in autism spectrum disorder, it can impact not only daily activities, but also social interactions [26]. However, one problem with implying predictive mechanisms in motor control is that prediction in real time is hard as agent-object interactions are not necessarily easy to predict – or not predictable at all.

Interestingly, recent studies have shown that humans simplify challenging control tasks by making their interactions more predictable. For example, when interacting with a double-pendulum exhibiting dynamics similar to a human limb, participants showed to be sensitive to resonance, even in such nonlinear systems, and exploit their inherent dynamics whenever possible [27]. While human daily activities involve manipulating various tools, ranging from tying shoelaces to drinking a glass of wine, only a limited number of studies have explored interactions with non-rigid objects [28–30]. Generalizing findings from traditional motor control studies to such scenarios is challenging. Interaction with complex dynamics requires strategies that do not rely on explicit knowledge of the nonlinear and potentially chaotic system dynamics. Yet, humans demonstrate remarkable dexterity in controlling and manipulating even the most complex underactuated objects—for example controlling a whip or carrying a cup of coffee without spilling.

The latter task has been studied in simplified form, simulated by a cup with a ball rolling inside. Findings revealed that humans prioritize predictability over conventional objectives, such as effort or smoothness. Specifically, they aim to make the human input to result in a more tractable and predictable output. One study by Sternad and collaborators [31] used mutual information between interactive force and cup-ball dynamics to quantify task predictability (Figure 1). They found that, over time, participants adopted control strategies with higher mutual information between the hand and the object, even at the expense of physical effort. Subsequent research [32] extended these findings to scenarios, where participants could freely choose the frequency of their interactions and leverage the mechanical impedance of their hands. Another study [33] demonstrated that participants, given the opportunity to set the system's initial conditions, such as the ball's angle and velocity, consistently converged to specific initial conditions that enhanced predictability and control. Experimental and modeling results also showed that subjects only needed to develop a simple and approximate representation of the cup-ball system's dynamics to achieve relative advanced tasks [34].

A whip, a flexible and underactuated object with theoretically infinite degrees of freedom, presents a similar, though vastly more difficult control challenge. Nonetheless, humans can reach astonishing skill at manipulating such objects. When attempting to hit a target with a whip, participants again were shown to adjust the initial conditions of their throw to simplify the system's behavior [9]. By orienting the whip backward and fully extending it before throwing, participants transformed the whip's motion into a simpler behavior, helping to achieve consistently good results. Despite controlling a whip is a redundant problem, we are able to learn control policies that select from many possible solutions (or initial states) the ones that render the task more predictable and controllable.

One final study on the control of an inverted pole suggests a link between increased predictability and control efficiency [35]. The most successful participants decreased the variability of the states they visited, at the expense of increased action variability. Such behavior suggests that participants sought solutions that were more stable – hence more predictable – and highlighted the key role of predictability in optimizing control strategies. However, while this study suggests that participants select predictable solutions, it does not directly test whether they actually predict the movements of the inverted pendulum. Future studies would be needed to test this hypothesis.

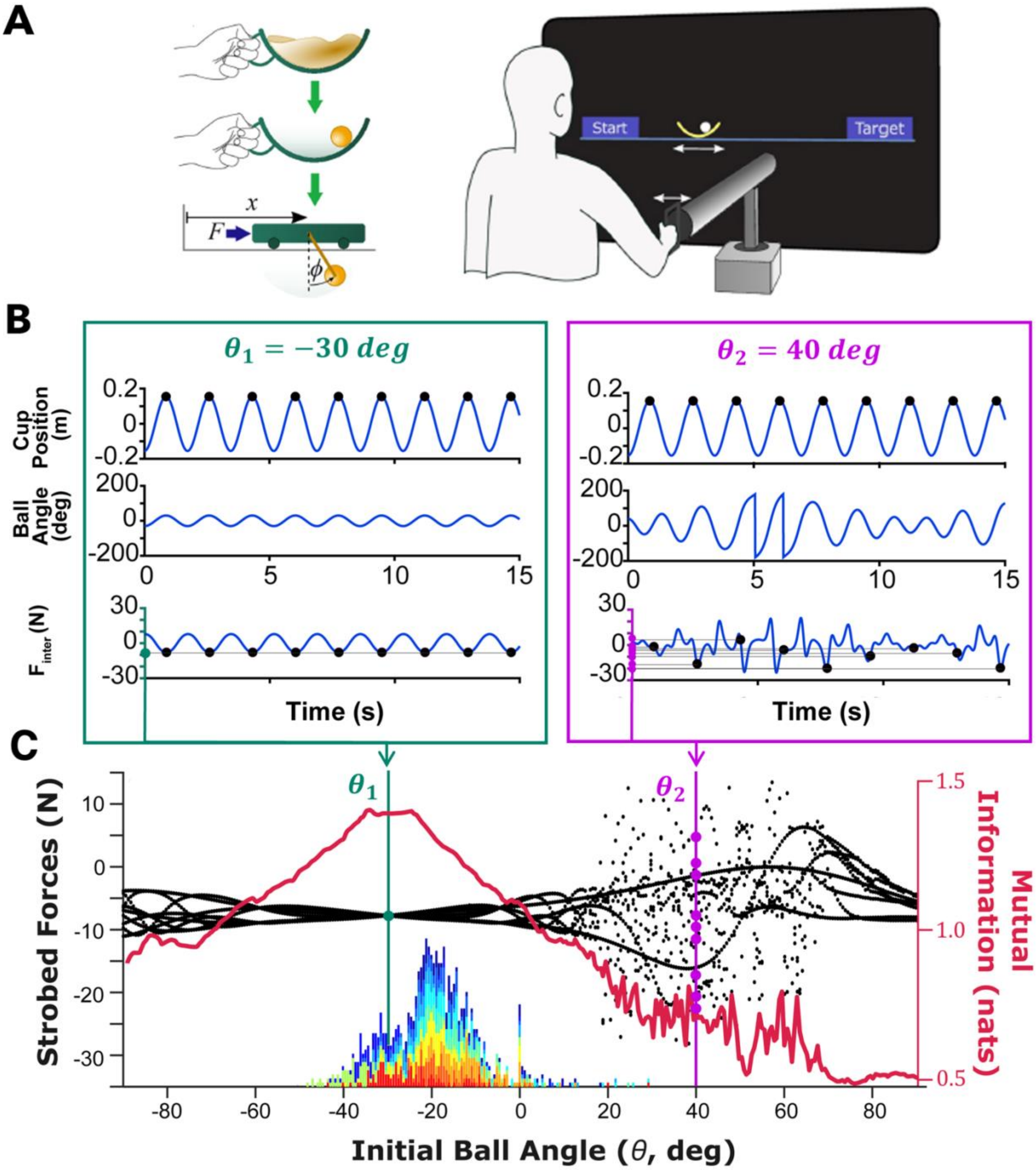

***Figure 1. Effect of Initial Ball Angle on Predictability of Interaction between the Human and the Object. (Panel A)** The task involved moving a two-dimensional cup with a ball sliding inside, inspired by humans carrying a cup of coffee. The system was mechanically equivalent to a cart sliding on a frictionless horizontal line with a suspended frictionless pendulum. A participant stands in front of the large screen and holds the handle of the HapticMaster robot (Moog FCS Robotics) to move the cup in the virtual environment. **(Panel B)** Exemplary simulated time series of the cup and ball and the associated interaction force $F_{inter}$ for two initial ball angles θ. $F_{inter}$ is strobed at the maxima of the sinusoidal cup position initiated with different initial ball angles (ranging from -90° to 90°). **(Panel C)** The marginal distributions obtained for different initial ball angles θ are plotted as a function of initial ball angle. For each theta, the black points are the marginal distributions of the strobed forces. While at angles between approximately 10° and 70°, there are complex distributions reflecting chaotic*

*force profiles, only a single force is found at -30°. A single force value corresponds to a sinusoidal interaction force, frequency-locked with the sinusoidal cup position. Mutual information (red) between the continuous interaction force and cup kinematics quantifies predictability for each initialization. Higher MI values signify greater predictability; they align closely with angles values that generate simpler force distributions. This match emphasizes the correlation between predictability (MI) and the complexity of the interaction force, with the highest MI values aligning with regions of stable and predictable dynamics. The histogram depicts the experimental distribution of initial ball angles pooled across all subjects (coded by color), coincident with those angles that generate a very regular - predictable – cup-force interactions (Figure modified with permission from [33]).*

**Predictability in joint action**

Humans excel in joint actions when collaborating with partners, such as rowing together, assembling furniture, or setting up tables for dinner, all requiring precise coordination and timing of one's actions in relation to the partners' actions [36–38].

Recent studies revealed that during joint action, we adopt a variety of different mechanisms [39,40]. These range from the synchronization of movements between co-actors, action predictions about what the partners are about to do, to higher-level mental state inference or "mentalizing" others' proximal goals, distal intentions and beliefs [41]. Recently, it has been demonstrated that synchrony during decision-making and movement phases enhances perceptions of cooperation within group activities [42].

Action prediction and mental state inference are challenging tasks, yet humans solve them efficiently and in real time. One line of research shows that when observing others' movements, we have access to their intentions through movement kinematics [43,44]. For example, we can infer the distal intention of a person grasping a bottle, e.g., pouring or drinking, by observing subtle changes in the grasping kinematics before the person touches the bottle [45]. Further, the kinematics of their reaching arm and hand preshape can also reveal the object that a person is about to grasp [46]. Recent evidence showed that humans can pick up and exploit a thrower's kinematic cues to increase the probability of correct ball interception [13]. And of course, expert throwers employ deceptive strategies to decouple the kinematics of their throwing action from the intended ball release [47].

These studies suggest that movement kinematics is a rich source of information, permitting the prediction of future actions and inference of proximal and distal intentions [48]. Recent technical development showed that the amount of information encoded in kinematics and its readout by observers can be precisely quantified [49], analogous to quantification of mutual information during hand-object interactions discussed above [31]. Recent works started investigating the neural underpinnings of interpersonal communication during joint actions. The intraparietal lobule was involved when reading out an observed action [50], and there was a synergistic coactivation of inhibitory and excitatory areas in the premotor cortex during the actual execution of joint actions [51,52].

Recent research revealed that movement characteristics can also indicate an individual's influence on others in a group setting, supporting the idea that movements serve as a valuable tool for examining how confidence is shared in collective human decision-making [53,54]. Interestingly, during joint action people do even more than reciprocal prediction and mental state inference: they continuously engage in non-verbal sensorimotor forms of communication to increase the information encoded in their movement kinematics. This in turn makes their behavior more predictable and their intentions easier to read by co-actors [55,56]. This is particularly the case under significant task uncertainty; for example, when one co-actor (the "leader") knows the task, but the other co-actor (the "follower") does not. Studies of leader-follower dynamics during a task where both simultaneously grasp two opposite sides of the same object show that leaders amplify their reaching gesture to distinguish it from alternative gestures, when the follower does not know the task [57,58]. This makes their behavior more legible and their action goals easier to infer by the follower [59,60].

Sensorimotor communication strategies can be very flexible as they are (also) modulated by task demands, such as the follower's uncertainty [61,62]. For example, co-actors instructed to coordinate key presses in a two-choice reaction task reduced their action variability compared to when they solve the task individually, plausibly as a way to increase their action predictability [63]. This is similar to linguistic communication, when speakers make their end of turn easy to predict [64]. The spatiotemporal alignment of movements during joint tasks can also serve as a way to increase the predictability of action timing [65]. Finally, sensorimotor communication can potentially convey various types of "messages" – illustrating that movement kinematics, or 'body language', manifests communicative intention [66]. Interestingly, both sensorimotor and linguistic communication can exploit predictive mechanisms: communicative messages are conveyed by "surprising" others, i.e., challenging their predictions in the short term, in order to update their shared understanding of the task in the long term [67–69].

**Predictive and non-predictive accounts – a continued discussion**

While the previous sections presented evidence for predictive mechanisms in single and joint actions, two methodological and theoretical issues need to be pointed out. First, it is important to distinguish between what appears to be predictable or predicted from an external observer's perspective and what a human may actually predict - or not. Even if a scientist identifies predictable regularities in a human's behavior, this does not automatically entail that the human is using prediction. The predictability that we observe or measure could simply be the outcome of a well-tuned control system [19]. Indeed, there is a close alignment between minimizing the discrepancy between predicted and sensed outcomes, as emphasized by predictive processing theories, and controlling perceptual inputs to reference states, as emphasized by non-predictive, closed-loop control theories. Interpreting empirical results requires utmost care to avoid premature conclusions.

Second, and relatedly, there are often multiple ways to solve a task, that may or may not require prediction. Predictive processing theories, active inference and optimal motor control suggest that people use forward models to make predictions of (for example) the ball trajectory for interception or of others' movements during joint actions, as in dancing together [3,16]. Possible theoretical advantages of making predictions include compensating for delays, especially in tasks like catching fast balls [10]. Another advantage is that prediction may optimize task-relevant aspects, such as end point accuracy [70] or, in dancing, smoothness of movement [23,71]. In contrast, non-predictive accounts, such as Perceptual Control Theory, argue that people might use memory and learning mechanisms - instead of prediction - to achieve complex sensorimotor tasks. For example, when controlling complex objects like a whip, they might learn which initial conditions are suitable for good results [1,2].

Another alternative to prediction in joint actions is that dancers may pursue common movement goals. Leaders provide signals about these common goals to followers, who understand them and move accordingly, ensuring that the two dancers synchronize [72]. Explanations based on predictive mechanisms and common goals are not necessarily mutually exclusive, since several predictive processing models use common goals to solve joint action tasks [69,73,74].

The debate between predictive and non-predictive accounts continues as seen in the example of interception and catching flyballs [10] [24]. Ultimately, the question of whether people use prediction is an empirical one. In this article, we highlighted several studies providing behavioral evidence for explicit predictions of moving objects during interception and avoidance tasks [23,75] and of other people's actions in joint action setups [11–15,41,68,76], but additional studies are needed to distinguish more clearly predictive versus non-predictive accounts of these phenomena.

## Conclusions

Humans exhibit exquisite motor skills, individually or together with others. Hierarchical, closed-loop control mechanisms in real time are fundamental to successfully perform these feats even in noisy and uncertain conditions [1,2]. Motivated by the recent framework of a "predictive brain", various lines of research suggest that predictive mechanisms supplement closed-loop control to solve both individual and joint motor control tasks. However, predicting object dynamics, anticipating others' actions and inferring their distal intentions is challenging. This leads to the fundamental question of *how* a "predictive brain" supports efficient individual and joint tasks. The studies reviewed here suggest a possible answer to this question: they show that we use "prediction tricks": we select subspaces of solution manifolds that afford predictable solutions and we make our sensorimotor interactions predictable for others. This might provide an explanation of how we solve these challenging tasks with apparent ease. Some of these ideas begin to be incorporated in artificial intelligence and in robotic applications [73,77]. These ideas also resonate with recent theoretical approaches that address complex skills, by assuming that they lie on low-dimensional manifolds. One example is a recent approach that successfully trained a musculoskeletal model on a complex object-manipulation task (rotating two Baoding balls in one hand) [78].

With its focus on "increasing predictability" – formalized by mutual information and stability in control strategies – this article aims to highlight the fundamental link between predictability, controllability, and behavioral efficiency. Increasing task predictability enhances closed-loop control by guiding behavior to subspaces of the solution manifold that afford accurate control towards the reference or goal, aligning efficiency in both prediction and control.

Given the ongoing discussion and remaining uncertainty about the underlying control processes, an important direction for future research consists in designing novel studies that carefully distinguish between predictive and non-predictive accounts – and to identify whether different contextual conditions, such as delays [23], might promote predictive or non-predictive mechanisms, even in the same tasks.

## Fundings

This research received funding from the European Union's Horizon 2020 Framework Programme for Research and Innovation under the Specific Grant Agreement No. 952215 (TAILOR); the European Research Council under the Grant Agreement No. 820213 (ThinkAhead), the Italian National Recovery and Resilience Plan (NRRP), M4C2, funded by the European Union – NextGenerationEU (Project IR0000011, CUP B51E22000150006, "EBRAINS-Italy"; Project PE0000013, "FAIR"; Project PE0000006, "MNESYS"), and the and the Ministry of University and Research, PRIN PNRR P20224FESY, PRIN 20229Z7M8N, and PRIN 2022YXLNR7_001. This research was also funded by National Institutes of Health, Grant Agreement No. NIH-R37-HD087089, NIH-R01-CRCNS-NS120579, National Science Foundation, Grant Agreement No. NSF-BCS-PAC-2043318, awarded to Dagmar Sternad.